\documentclass[a4paper,11pt]{article}
\pdfoutput=1

\usepackage{jheppub}

\usepackage[T1]{fontenc}
\usepackage{caption}
\usepackage{subcaption}
\usepackage{dsfont}
\usepackage{extarrows}
\usepackage[aligntableaux=center,boxsize=0.8em]{ytableau}

\allowdisplaybreaks[1]

\def\({\left(}
\def\){\right)}
\def\[{\left[}
\def\]{\right]}
\def\<{\left\langle}
\def\>{\right\rangle}

\def\be{\begin{eqnarray}}
\def\ee{\end{eqnarray}}
\def\nn{\nonumber\\}
\def\pa{\partial}

\def\zb{\bar{z}}

\def\g{\gamma}
\def\G{\Gamma}
\def\d{\delta}
\def\D{\Delta}
\def\e{\epsilon}

\def\l{\lambda}

\def\s{\sigma}

\def\t{\tau}

\def\nn{\nonumber\\}
\def\pa{\partial}

\title{Easy bootstrap for the 3D Ising model: \\
a hybrid approach of the lightcone bootstrap and error minimization methods}

\author{Wenliang Li}
\affiliation{School of Physics, Sun Yat-Sen University, Guangzhou 510275, China}

\emailAdd{liwliang3@mail.sysu.edu.cn}

\abstract{As a simple lattice model that exhibits a phase transition, 
the Ising model plays a fundamental role in statistical and condensed matter physics.   
The Ising transition is realized by physical systems, such as the liquid-vapor transition. 
Its continuum limit also furnishes a basic example of interacting quantum field theories 
and universality classes.   
Motivated by a recent hybrid bootstrap study of the quantum quartic oscillator, 
we revisit the conformal bootstrap approach to the 3D Ising model at criticality, 
without resorting to positivity constraints. 
We use at most 10 nonperturbative crossing constraints at low derivatives  
from the Taylor expansion around a crossing symmetric point.  
The high-lying contributions are approximated by simple analytic formulae 
deduced from the lightcone singularity structure.  
Surprisingly, the low-lying properties are determined to good accuracy by  
this computationally very cheap approach. 
For instance, the results for the two relevant scaling dimensions 
$(\D_\s,\D_\e)\approx (0.518153,1.41278)$ are close to the most precise rigorous bounds 
obtained at a much higher computational cost. 
}

\begin{document} 
\maketitle
\flushbottom
\section{Introduction}
Onsager's exact solution for the partition function of the 2D square-lattice Ising model 
is a landmark in theoretical physics \cite{Onsager:1943jn}. 
However, an exact solution for the 3D cubic-lattice Ising partition function  
has remained elusive. 
Since the 2D Ising model is exactly solvable, 
one may try to embed the simple cubic lattice
in a non-planar surface.   
This is indeed possible \cite{Regge:1999bg}, 
but the price to pay is that the genus is proportional to the number of sites,  
which becomes infinite in the thermodynamic limit. 
Then the partition function is given by an infinite sum of complicated terms.   
\footnote{The partition function is given by a sum of $4^{g}$ Pfaffians \cite{Kasteleyn}, where $g$ is the topological genus and $4^{g}$ is the number of spin structures.}
The 2D embedding does not seem to be particularly useful 
due to the interplay between the dimensional complexity and topological complexity.  
This also explains why it should be significantly more difficult to solve the Ising model in $D>2$ dimensions. 

Although an exact solution at $D>2$ may be challenging, 
the long distance behavior of the Ising model is not so complex. 
\footnote{It is interesting to study the $D$-dimensional Ising model 
using the $\epsilon=D-2$ expansion. 
See \cite{Li:2021uki} for a conformal bootstrap study of the $D=2+\e$ Ising model at criticality.} 
There is an emergent simplicity in the macroscopic physics. 
For example, the scaling behavior around the critical point is 
characterized by some critical exponents, 
which can be determined to good precision by the Monte Carlo simulations of finite size systems 
(see e.g. \cite{Hasenbusch:2010hkh, Hasenbusch:2021tei} and references therein). 
To study the macroscopic properties, 
one can further consider the continuum limit,  
whose effective description is expected to be given by the $\phi^4$ field theory. 
The long stance behavior is determined by an infrared fixed point of the renormalization group (RG) flow, 
corresponding to a scale invariant quantum field theory.
Different physical systems can exhibit similar macroscopic properties 
if they are governed by the same fixed point, i.e.,  in the same universality class. 
In $D>4$ dimensions, the Ising critical exponents can be derived from the simple mean field theory 
due to the averaging effects of a large number of nearby sites, 
associated with the Gaussian fixed point of the RG flow. 
For $1<D<4$, the Ising critical behavior is governed by a nontrivial fixed point of the RG flow 
and the critical exponents become nontrivial functions of $D$. 
Furthermore, there is strong numerical evidence that 
scale invariance of the 3D Ising fixed point is enhanced to conformal invariance
\cite{El-Showk:2012cjh,El-Showk:2014dwa,Kos:2014bka,Simmons-Duffin:2015qma,Kos:2016ysd,Billo:2013jda,Cosme:2015cxa,Zhu:2022gjc,Hu:2023xak,Lao:2023zis,Han:2023yyb, Hu:2023ghk}. 
It should be easier to solve the 3D Ising model at criticality. 

Since the revival of  the $D>2$ conformal bootstrap program \cite{Rattazzi:2008pe}, 
one of the most impressive achievements is the high precision determinations of 
the low-lying data of the 3D Ising conformal field theory (CFT) 
\cite{El-Showk:2012cjh,El-Showk:2014dwa,Kos:2014bka,Simmons-Duffin:2015qma,Kos:2016ysd}. 
\footnote{In $D=2$ dimensions, the Ising conformal field theory is exactly solvable by the conformal bootstrap method and is associated with the simplest unitary minimal model \cite{Belavin:1984vu}. } 
The precise scaling dimensions and operator-product-expansion (OPE) coefficients of the two relevant operators are \cite{Kos:2016ysd}
\be
\D_\s=0.5181489(10)\,,\quad
\D_\e=1.412625(10)\,, \quad
\l_{\s\s\e}=1.0518537(41)\,.
\label{island-results}
\ee
The determinations of critical exponents
$\eta=0.0362978(20)$, $\nu=0.629971(4)$ 
are consistent with and more precise than the Monte-Carlo determinations 
$\eta^\text{MC}=0.036284(40)$, 
$\nu^\text{MC}=0.62998(5)$ in \cite{Hasenbusch:2021tei}. 

In \eqref{island-results}, the error bars from the numerical bootstrap bounds are rigorous. 
The tiny allowed region is deduced from 
the assumptions of unitarity, two relevant operators, and OPE associativity.  
The unitarity assumption implies that the OPE coefficients are real numbers,  
leading to positivity constraints.  
The number of relevant operators and the unitarity bounds give 
some lower bounds for the scaling dimensions of the irrelevant operators. 
\footnote{In fact, spectral boundedness is a strong assumption \cite{Li:2022prn,Li:2023nip}, 
which can play a similar central role in the nonpositive bootstrap as
the positivity constraints in the bootstrap bound approach. }
For 4-point functions, the associativity of OPE leads to the crossing equations, 
i.e., 
the convergent summations of conformal blocks in different channels should correspond to the same correlator. 
The crossing equations can be discretized by the Taylor expansion 
around a well-chosen point in the rapidly convergent regime, 
where the dominant contributions are associated with operators of low scaling dimension \cite{Pappadopulo:2012jk}. 
These nonperturbative constraints are labelled by the numbers of derivatives. 
Together with the positivity constraints and spectral assumptions, 
one can use the crossing constraints to rule out the inconsistent parameter space. 
Using a system of mixed correlators, an isolated region of the parameter space 
for the 3D Ising CFT was found in \cite{Kos:2014bka}.

If a correlator admits a rapidly convergent expansion in terms of conformal blocks, 
the high-lying contributions are suppressed and thus less constrained by the crossing equation. 
On the other hand, it is mathematically more subtle to study the implications of 
the crossing equation in a singular limit, 
as the conformal block summation may cease to be manifestly convergent. 
\footnote{See the section 2.4 of \cite{Penedones:2019tng} for some examples of dangerous limits. }
Nevertheless, a singular limit of the crossing equation can provide nontrivial information 
about the high-lying states, which is complementary to the rapidly convergent case.   
The reason is that the emergence of a singularity requires a collective behavior of infinitely many high-lying states,  
so one may obtain simple asymptotic formulae for the averaged behavior at large quantum numbers. 
For example, in the lightcone limit, 
the dominant contribution of the cross-channel vacuum state 
implies the double-twist behavior at large spin \cite{Fitzpatrick:2012yx,Komargodski:2012ek}.  
\footnote{We assume that the identity operator has the lowest twist. 
In perturbation theory, the twist accumulation phenomenon was discussed already in 1973 \cite{Parisi:1973xn, Callan:1973pu}. 
See also \cite{Alday:2007mf} for an earlier nonperturbative argument based on a 2D massive theory perspective.  
Recently, a more rigorous approach was presented in \cite{Pal:2022vqc} based on positivity constraints.}
The leading asymptotic behavior coincides with that of the generalized free fields. 
At low twist, the higher spin results from the numerical bootstrap bounds 
and the extremal functional method \cite{Poland:2010wg, El-Showk:2012vjm, El-Showk:2014dwa, Simmons-Duffin:2016wlq} 
beautifully organize into Regge trajectories, 
in accordance with the analytic formulae from the lightcone bootstrap 
\cite{Alday:2015ota, Alday:2015ewa, Simmons-Duffin:2016wlq}. 

The high precision numerical bootstrap bounds \eqref{island-results}  
are obtained from expensive computations. 
Suppose that we make use of a global understanding of the high-lying operators.  
{\it Can we determine the low-lying properties of the 3D Ising CFT reasonably well 
with less computational efforts?} 
Recently, we performed a bootstrap study of the quartic anharmonic oscillator \cite{Bender:1969si} 
using the asymptotic behavior and nonperturbative self-consistency constraints 
\cite{Li:2023ewe}, 
which can be viewed as a bootstrap study of the 1D version of the massive $\phi^4$ field theory.
\footnote{This study was inspired by the recent work on the 0D $\phi^4$ theory \cite{Bender:2022eze,Bender:2023ttu} 
and a bootstrap formulation of the quantum anharmonic oscillator \cite{Han:2020bkb}. 
See \cite{Li:2024rod} for more details about a novel type of analytic continuation. }
As the 1D bootstrap results exhibit rapid convergence, 
it is tempting to apply this strategy to the 3D field theory. 

In \cite{Simmons-Duffin:2016wlq}, Simmons-Duffin suggested a hybrid bootstrap  approach 
that combines the lightcone approach \cite{Fitzpatrick:2012yx,Komargodski:2012ek} with the truncation approach \cite{Gliozzi:2013ysa}. 
\footnote{See \cite{Iliesiu:2018zlz} for a study of the Ising CFT at finite temperature using this hybrid approach. }
In the truncation method initiated by Gliozzi, 
the contributions of the high-lying operators are simply set to zero. 
As the high-lying contributions are suppressed near $z=\zb=1/2$,  
it seems reasonable to omit them in a preliminary study. 
However, if one wants to systematically improve the results, 
some small but non-negligible high-lying contributions should be taken into account. 
To determine the additional free parameters associated with the high-lying contributions, 
one needs to consider more nonperturbative crossing constraints. 
Without the help of positivity \footnote{See \cite{Su:2022xnj} for a different hybrid approach that combines the positivity constraints with the large spin perturbation theory. }, 
the computational complexity of a large system of nonlinear equations grows rapidly. 
\footnote{Using the rational approximations \cite{Kos:2013tga}, 
the transcendental equations can be well approximated by polynomial equations, 
which can be studied by numerical algebraic geometry methods, 
such as the efficient package  \texttt{HomotopyContinuation.jl} \cite{Breiding-Timme}. 
However, the computational complexity still grows rather rapidly. }
In a hybrid approach, 
the high-lying scaling dimensions and OPE coefficients are approximated by 
simple analytic formulae from the lightcone bootstrap. 
As the number of free parameters is reduced, 
less crossing constraints are needed.  
In this way, one should be able to improve the results more easily. 
In \cite{Simmons-Duffin:2016wlq}, the original proposal involves the crossing constraints for mixed correlators. 
Below, we will show that the hybrid analytical/numerical approach can be realized in a simpler way, 
i.e., by studying only an identical correlator. 
\footnote{The reason is that the mixing of subleading Regge trajectories does not have dominant effects on the identical correlator around $z=\bar z=1/2$. 
A CFT with large mixing effects may not be captured by the crossing equation for an identical correlator.  }
As discussed later, we are led to an error minimization approach \cite{Li:2017ukc} 
in the hybrid method.

In this work, we revisit the conformal bootstrap approach to the 3D Ising model 
using the two complementary types of crossing constraints:
\begin{itemize}
\item
The singular $\bar z\rightarrow 1$ limit leads to a global understanding of the high-lying data.  
\item
The expansion around
$z=\zb=1/2$ gives strong constraints on the low-lying data. 
\end{itemize}
We assume that the 3D Ising CFT has only two relevant operators.  
We will consider the 4-point correlator of the lowest $\mathbb Z_2$-odd scalar $\s$ with $\D_\s<3$. 
The other relevant operator $\e$ is a $\mathbb Z_2$-even scalar and appears in the $\s\times\s$ OPE. 
It turns out that reasonably accurate results can be derived easily 
from a small number of nonperturbative crossing constraints. 
In our main results, the total number of derivatives in the expansion around $z=\zb=1/2$ is at most $7$.

This paper is organized as follows. 
In Sec. \ref{Sec_crossing}, we give a brief overview of the crossing constraints. 
In Sec. \ref{Sec_I}, we discuss how to obtain an approximate solution of the 3D Ising CFT using simple analytic approximations and a few nonperturbative crossing constraints. 
In Sec. \ref{Sec_low-lying}, 
we take into account the leading corrections to the analytic approximations for the high-lying contributions. 
We use the error minimization approach \cite{Li:2017ukc} 
to obtain more accurate results of the 3D Ising CFT data. 
In Sec. \ref{Sec_discussion}, we summarize our results and discuss some directions for further investigations. 

\section{Crossing constraints}
\label{Sec_crossing}
We will consider the 4-point correlator of the lowest $\mathbb Z_2$-odd scalar $\s$:
\be
\langle \s(x_1)\s(x_2)\s(x_3)\s(x_4)\rangle
=\frac{G(z,\zb)}{x_{12}^{2\D_\s}x_{34}^{2\D_\s}}\,,
\ee
where $x_i$ denote the positions of the external operators, 
and $x_{ij}=|x_i-x_j|$ are the distances between two operators. 
The variables $(z, \zb)$ are defined by
\be
z\zb=\frac{x_{12}^2x_{34}^2}{x_{13}^2x_{24}^2}\,,\quad
(1-z)(1-\zb)=\frac{x_{14}^2x_{32}^2}{x_{13}^2x_{24}^2}\,.
\ee
The correlator is invariant under the exchange of $\s(x_1)$ and $\s(x_3)$, 
so we have the crossing equation
\be
(1-z)^{\D_\s} (1-\zb)^{\D_\s} G(z,\bar z)=
(z\bar z)^{\D_\s} G(1-z,1-\bar z)\,,
\label{crossing}
\ee
which gives a relation between the direct-channel and cross-channel OPEs
 and the corresponding conformal block expansions. 
We can discretize the crossing equation \eqref{crossing}
by the Taylor expansion around $z=\zb=1/2$:
\be
\sum_{i}P_i\,\mathcal F^{(m,n)}_{\D_i,\ell_i}=0\,,
\label{crossing-derivatives}
\ee
where $P_i=\l^2_{\s\s\mathcal O_i}$ are the squared OPE coefficients, and
$(\D_i,\ell_i)=(\D_{\mathcal O_i},\ell_{\mathcal O_i})$ are the scaling dimension and spin of the primary operator $\mathcal O_i$. 
The definition of $\mathcal F^{(m,n)}_{\D_i,\ell_i}$ is
\footnote{In this work, the conformal blocks are computed numerically using the \texttt{simpleboot} package \cite{simpleboot}.}
\be
\mathcal F^{(m,n)}_{\D_i,\ell_i}=
\pa_z^m \pa_{\zb}^n\big[(1-z)^{\D_\s} (1-\zb)^{\D_\s} F_{\D_i,\ell_i}(z,\bar z)
-(z\leftrightarrow 1-\zb)\big]
\big|_{z=\zb=\frac 12}\,.
\ee
By construction, the nonperturbative crossing constraints \eqref{crossing-derivatives} vanish automatically if $m+n$ is an even integer. 
Since the crossing equation \eqref{crossing} is symmetric in $(z,\zb)$, 
we assume $m<n$. 
In this work, the normalization of a conformal block is fixed by 
\be
F_{\D,\ell}(z,\zb)=z^{(\D-\ell)/2}\zb^{(\D+\ell)/2}(1+\mathcal O(\zb))+\mathcal O(z^{(\D-\ell)/2+1})\,,
\ee
where $\D$ is the scaling dimension and $\ell$ is the spin. 

As explained in the introduction, 
it is also useful to consider the crossing equation in the singular lightcone limit $\zb\rightarrow 1$. 
After rewriting the crossing equation \eqref{crossing} as
\be
\sum_{i}P_i\, F_{\D_i,\ell_i}(z,\zb)=
\frac{(z\bar z)^{\D_\s}}{(1-z)^{\D_\s} (1-\zb)^{\D_\s} } \sum_{i} P_i\, F_{\D_i,\ell_i}(1-\zb,1-z)\,,
\label{lightcone-crossing}
\ee
one can show that the leading asymptotic behavior at large spin is 
associated with the lowest twist operator on the right hand side. 
The asymptotic behavior can be derived from the relation 
\be
\sum_{h-h_0=0,2,4,\dots}2S_p(h)\, k_h(\zb)\sim \Big(\frac {1-\zb}{\zb}\Big)^p
\quad (\zb\rightarrow 1)\,,
\label{asymptotic-sum}
\ee
where the coefficient function $S_p(h)$ is
\be
S_p(h)\equiv\frac{\G(h-p-1)}{\G(-p)^2\,\G(h+p+1)}\frac {\G(h)^2}{\G(2h-1)}\,,
\ee
the $SL(2,R)$ block $k_h(\zb)$ is 
\be
k_h(\zb)\equiv\zb^h{}_2F_1(h,h,2h,\zb)\,,
\ee
and the conformal spin $h$ is related to the twist $\t$ and spin $\ell$ by
\be
h\equiv\frac \t 2+\ell\,.
\ee 
An exact identity version of \eqref{asymptotic-sum} can be found in \cite{Simmons-Duffin:2016wlq}.

\section{Asymptotic behavior from the vacuum state}
\label{Sec_I}
The leading asymptotic behavior is associated with the vacuum state in the cross-channel, 
as the identity operator has the lowest twist in the 3D Ising CFT. 
Using \eqref{asymptotic-sum} and the small $z$ expansion of the conformal block $F_{\D,\ell}(z,\zb)$, 
one can derive the large-spin asymptotic behavior of the squared OPE coefficients
\be
P_{\{I\}}^{(k=0)}(h)\sim 2S_{-\D_\s}(h)\,,
\label{P-I-0}
\ee
\be
P_{\{I\}}^{(k=1)}(h)\sim \frac{(\D_\s-D/2+1)(h-\D_\s)(h+\D_\s-1)}{(h-\D_\s+D/2-1)(h+\D_\s-D/2)}S_{-\D_\s}(h)\,,
\label{P-I-1}
\ee
where the superscript $(k)$ denotes the trajectory $[\s\s]_{k,\ell}$ with twist $\t\sim 2\D_\s+2k$.  
The subscript $\{I\}$ indicates that the asymptotic behavior is associated with the contribution of the cross-channel vacuum state.   
The analytic formulae \eqref{P-I-0},\eqref{P-I-1} coincide with the OPE coefficients of the generalized free fields (GFF)
\cite{Heemskerk:2009pn,Fitzpatrick:2011dm}. 
They provide a simple approximation for the averaged behavior of the high-lying states. 
In fact, the scaling dimensions deviate from the double-twist values $\D_{[\s\s]_{k,\ell}}=2\D_\s+2k+\ell$ at finite spin, 
but the effects of their corrections are subleading. 
Some of these corrections will be taken into account to improve the accuracy 
in Sec. \ref{Sec_low-lying}. 

The low-lying properties deviate more significantly from the GFF solution,  
so we should separate the low-lying contributions from the high-lying ones.  
The approximate $G(z,\bar z)$ contains a low-lying part and a GFF part
\be
G(z,\bar z)&\approx& 1+P_\e F_{\D_\e,0}(z,\bar z)+P_T F_{3,2}(z,\zb)+\text{(GFF)}\,,
\label{G-I}
\ee
where the GFF part does not include the cases of $\t=2\D_\s$ with $\ell=0,2$. 
Here $(\D_\e, P_\e, P_T)$ are free parameters. 
Note that $\e$ is the $\mathbb Z_2$-even relevant operator with $0<\D_\e<D$. 
We assume that the lowest spin-two operator corresponds to the stress tensor $T$, 
so its scaling dimension is given by $\D_T=D=3$. 
As the generalized free fields correspond to an exact solution to the crossing equation, 
we can sum over the GFF contributions and 
obtain a more explicit form of  \eqref{G-I}: 
\be
G(z,\zb)&\approx&
1+(z\zb)^{\D_\s}+\frac {(z\zb)^{\D_\s}}{[(1-z)(1-\zb)]^{\D_\s}}
\nn&&+P_\e F_{\D_\e,0}(z,\zb)+P_T F_{3,2}(z,\zb)-\sum_{\ell=0,2}2S_{-\D_\s}(\D_\s+\ell)F_{2\D_\s+\ell,\ell}(z,\zb)\,.
\ee
The first line contains the leading terms in the lightcone limit,  
which is manifestly crossing symmetric and thus reminiscent of the inverse bootstrap approach \cite{Li:2017agi}.  
\footnote{See \cite{Padayasi:2023hpd} for a recent application of the inverse bootstrap approach to 
the $D>2$ Anderson transitions.}
The second line is subleading in both the direct-channel and cross-channel lightcone limits, 
but becomes important around $z=\zb=1/2$. 
To satisfy the crossing constraints \eqref{crossing-derivatives}, 
the contributions of the low-lying operator $\{\e, T\}$ should satisfy 
\be
P_\e\, \mathcal F^{(m,n)}_{\D_\e,0}+P_T\, \mathcal F^{(m,n)}_{3,2}
-
\sum_{l=0,2}2S_{-\D_\s}(\D_\s+\ell)\, \mathcal F^{(m,n)}_{2\D_\s+\ell,\ell}
\approx 0 \,, 
\label{crossing-I}
\ee 
which leads to nontrivial constraints on the free parameters $(\D_\e, P_\e, P_T)$ and $\D_\s$. 
For comparison, we also examine the naive truncation without any high-lying contribution.  
The corresponding  crossing constraints read
\be
\text{(naive truncation)}\quad \mathcal F^{(m,n)}_{0,0}+P_\e\, \mathcal F^{(m,n)}_{\D_\e,0}+P_T\, \mathcal F^{(m,n)}_{3,2}\approx 0\,,
\text{}
\label{naive-truncation} 
\ee
which contain no contributions from primary operators with $\D>3$. 
The low derivative equations are more sensitive to the low-lying contributions due to the rapid convergence in the Euclidean regime \cite{Pappadopulo:2012jk}, 
so they are expected to be more useful for constraining the low-lying CFT data. 
\footnote{For the rapid OPE convergence, 
it is natural to consider the crossing symmetric point $z=\bar z=1/2$ as the reference point  
because we want to minimize the radial distance \cite{Pappadopulo:2012jk} 
in both the direct- and cross- channels. 
The region around $z=\bar z=1/2$ is still expected to be rapidly convergent in both channels
and they can be well approximated by the leading terms of the Taylor series. 
The low derivative crossing constraints can be viewed as certain linear combinations of 
those around $z=\bar z=1/2$, so they are expected to rapidly convergent as well. 
On the other hand, 
we need to use many higher derivative terms in the Taylor expansion to approach the singular lightcone limit 
and many higher spin operators to reconstruct the singularity structure. 
Therefore, we expect that the high derivative crossing constraints are more sensitive to the high-lying contributions than the low derivative ones. 
} 
According to the total number of derivatives, 
the leading nonperturbative equations correspond to
\be
(m,n)=(0,1),(0,3),(1,2)\,.
\label{derivative-I}
\ee 
For a given $\D_\s$, 
we can determine $(\D_\e,\,P_\e,\, P_T)$ by these three equations.

\begin{figure}[h]
	\centering
		\includegraphics[width=1\linewidth]{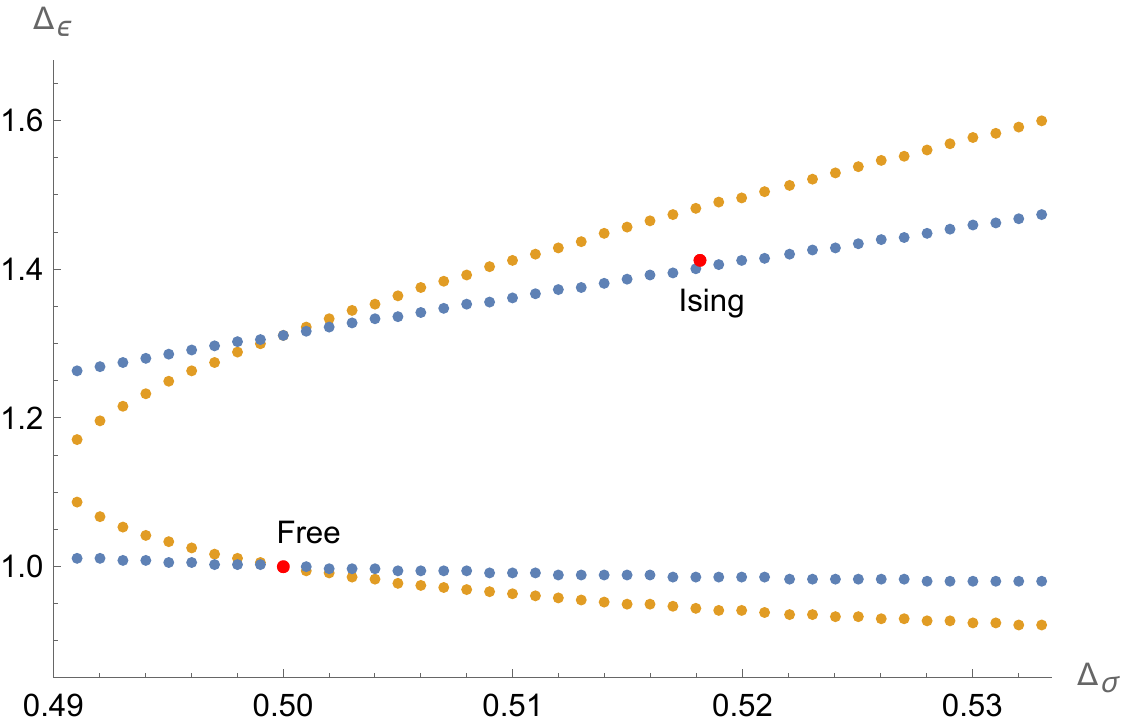}
	\caption{The approximate solution for $\D_\e$ at various $\D_\s$. The red dots represent the free theory and the Ising solutions in three dimensions. 
	The blue dots are obtained from solving \eqref{crossing-I} with GFF approximations for the high-lying contributions. 
	The orange dots are derived from solving \eqref{naive-truncation} with the naive truncation 
	that omits all the high-lying contributions. 
	The nonperturbative crossing constraints are associated with the derivative orders in \eqref{derivative-I}. 
	The results with the GFF approximations (blue) are much closer to the 3D Ising solution \eqref{island-results}. 
}
	\label{I-plot}
\end{figure}

In Fig. \ref{I-plot}, we present the solution for $\D_\e$ at various $\D_\s$. 
The blue dots denote the solutions with GFF approximations for the high-lying contributions, 
while the orange dots are associated with the naive truncation without any $\D>3$ contribution.  
We can see that the two types of solutions intersect around the free theory solution, 
but the results with GFF approximations are much closer to the 3D Ising solution \eqref{island-results}. 
\footnote{Note added: The improvements by GFF or mean field approximations are also reported in \cite{Poland:2023bny}, which further applies this approach to the five-point bootstrap. 
In \cite{Poland:2023bny}, the scaling dimensions are input parameters from the positivity bounds and the GFF approximations are used to improve the determinations of the OPE coefficients. 
Here we use the GFF improved truncation method to determine both the scaling dimensions and OPE coefficients without relying on any input parameters from the positivity bounds. 
It is more nontrivial to determine the scaling dimensions than only the OPE coefficients. 
In Sec. \ref{Sec_low-lying}, we further consider the leading corrections to the GFF approximations, which are not taken into account in \cite{Poland:2023bny}.  
Another difference is that the best known value of $\D_\s$ is used to select the set of nonperturbative crossing constraints by viewing certain $\D_\s$ in the crossing constraints as a free parameter 
in \cite{Poland:2023bny}.  }

To determine the external scaling dimension $\D_\s$, we should introduce one more equation. 
If we choose $(m,n)=(1,4)$, the solutions for the free parameters are
\be
\D_\s\approx 0.5177\,,\quad
\D_\e\approx 1.399\,,\quad
P_\e\approx 1.113\,,\quad
P_T\approx 0.1065\,,
\label{sol-GFF}
\ee
where we only write 4 significant digits. 
Despite the crude approximation scheme, 
the solutions of the crossing constraints are unexpectedly close to the correct values in \eqref{island-results} and \cite{Simmons-Duffin:2016wlq}. 
The relative errors are $(-0.08\%,\, -0.9\%, \,0.6\%, \,0.1\%)$. 
\footnote{
For spinning operators of low twist, it may be more appropriate to consider a percentage based error in terms of twist. }
These results are more accurate than those in the previous 3D Ising studies \cite{Gliozzi:2013ysa, Gliozzi:2014jsa} by Gliozzi's truncation method, which considered more scaling dimensions as free parameters and 
relied on some external input from the Monte Carlo method. 
The only exception is the prediction for $\D_\e$ in the most complicated example in \cite{Gliozzi:2014jsa} with 10 free parameters for the scaling dimensions and 1 input scaling dimension from the Monte Carlo method.
Let us emphasize that we do not use any external input for the scaling dimensions in this work. 
If all the scaling dimensions are fixed, it is trivial to determine the OPE coefficients 
by the error minimization method \cite{Li:2017ukc}. 

We can also estimate the errors without using the results from the positivity bounds. 
However, it is not clear how to perform a rigorous error estimation in the truncation-like method, 
as inequality-like positivity constraints are not used. 
Here we follow a preliminary procedure that 
does not allow for correlated movement in the variables. 
Let us consider the remaining 5-derivative crossing constraints with $(m,n)=(0,5),(2,3)$.  
We substitute the free parameters with the crossing solutions in \eqref{sol-GFF}, 
except for the one under error estimation. 
Since we use two constraints, we obtain two different values for the parameter under error estimation. 
The error is set by the largest deviation of this parameter from the crossing constraints.  
The results are
$
|\d\D_\s|=0.0084\,,
|\d\D_\e|=0.027\,,
|\d P_\e|=0.029\,,
|\d P_T|=0.0019$,
which are greater than the error estimates based on the positivity bound results. 
The error estimation method is similar to the one in \cite{Li:2017ukc}. 
Here we can also introduce the $\eta$ function and solve the chosen constraints by the $\eta$ minimization. 
Since we can solve all the constraints, the corresponding local minimum of the $\eta$ function is zero  
and we are not able to use the four constraints to estimate the errors. 
Therefore, we use the two 5-derivative equations to compute the errors. 
As the two constraints are not encoded in the $\eta$ function, 
the resulting errors are much greater than the estimates based on the positivity bounds. 

The critical exponents are related to the scaling dimensions by 
$\D_\s=1/2+\eta/2$, $\D_\e=3-1/\nu$, so we have
\be
\eta\approx 0.0355\,,\quad\nu\approx 0.6247\,,
\ee
where the relative errors of the crossing constraint solutions are also small, i.e. $(-2\%,0.8\%)$. 
The error estimates from the crossing constraints with $(m,n)=(0,5),(2,3)$ are
$|\d\eta|=0.0101$ and $|\d\nu|=0.0168$.

If we use a different 5-derivative equation, the results would not be so encouraging. 
For the $(0,5)$ and $(2,3)$ constraints, 
the solution in the upper blue branch in Fig. \ref{I-plot} becomes $\D_\s\approx 0.504$ and $\D_\s\approx 0.494$ respectively. 
\footnote{For the naive truncation \eqref{naive-truncation},  
we do not find a reasonable solution for the Ising scaling dimension $\D_\s$ using a five-derivative equation. }
One may suspect that the nice results in \eqref{sol-GFF} are a pure coincidence.  
To address this concern, 
we will take into account the leading corrections to the GFF behavior in Sec. \ref{Sec_low-lying}, 
and show that the results can be further improved.  
Therefore, the encouraging results \eqref{sol-GFF} indeed suggest a cheaper avenue towards 
an accurate solution of the 3D Ising CFT. 

\section{Asymptotic behavior from the low-lying states}
\label{Sec_low-lying}
To obtain more accurate results, we should improve the approximation scheme for the high-lying contributions. 
For the lowest $\mathbb Z_2$-even Regge trajectory with $\t\approx 2\D_\s$, 
the leading corrections to the GFF behavior can be easily deduced from the cross-channel operators $\{\e, T\}$, 
which will be discussed later.  
For the higher Regge trajectories, 
the GFF asymptotic behavior is an approximation for the averaged contributions.
\footnote{The higher trajectories are expected to exhibit more complex behaviors 
due to the mixing and exponentiation effects. 
}
We need to study a mixed system of correlators to extract the unmixed trajectories. 
Above the lowest trajectory $[\s\s]_{0,\ell}$, there are two dominant trajectories with $\t_{[\s\s]_{1}}\sim 2\D_\s+2$ and $\t_{[\e\e]_{0}}\sim 2\D_\e$.  
Their twist spectra are nearly symmetric about $\t=\D_\s+\D_\e+1$
and have large repulsion at low conformal spin. 
As the twists of the $\ell=0,2$ operators have significant deviations from the GFF value $\t=2\D_\s+2$, 
we will consider their twists as free parameters. 
Nevertheless, 
their squared OPE coefficients can be approximated by the GFF asymptotic formula \eqref{P-I-1}. 
This is because the $\ell=0,2$ operators on the lower trajectory $[\e\e]_{0,\ell}$ decouple 
\footnote{We believe that the absence of these two operators is closely related to 
the leading kink behavior in \cite{El-Showk:2012cjh,El-Showk:2014dwa}. 
They may violate the positivity constraints if present \cite{Li:2020ijq}.  }
and only one dominant operator contributes at each spin $\ell=0,2$.  
At higher spin or higher twist, we will again use the simple GFF approximation 
as the corrections are highly suppressed by the small OPE coefficients.

As mentioned above, we should take into account the leading corrections to the GFF approximation 
for the lowest trajectory $[\s\s]_{0,\ell}$, 
which can be deduced from the cross-channel contributions of the low-lying operators $\{\e, T\}$. 
In the lightcone limit $\zb\rightarrow 1$, the leading terms of the cross-channel conformal blocks are
\be
F_{\D,\ell}(1-\zb,1-z)&=&(1-\zb)^{\t/2}f_{h}(1-z)+\mathcal O[(1-\zb)^{\t/2+1}]
\nn&=&-\frac{\G(2h)}{\G(h)^2}(1-\zb)^{\t/2}\big(2H_{h-1}+\log z\big)
+\dots\,,
\ee
where $H_x=\G'(x+1)/\G(x+1)-\G'(1)/\G(1)$ denotes the harmonic number and $\dots$ indicates subleading terms in $1-\bar z$ or $z$. 
The $\log z$ terms are related to the anomalous dimensions by 
$z^{\t/2}=z^{\D_\s}(1+\frac {\t-2\D_\s} {2}\log z+\dots)$. 
The crossing equation \eqref{lightcone-crossing} implies 
that the leading behavior of the anomalous dimensions 
and the OPE coefficients are 
\be
\g_{\{I,\e,T\}}^{(k=0)}(h)\sim  \frac{-4\sum_{i=\e,T}\tilde P_i\, S_{-\D_\s+\t_i/2}(h)
}{P_{\{I,\e,T\}}^{(k=0)}(h)}\,,
\ee
\be
P_{\{I,\e,T\}}^{(k=0)}(h)\sim 
2S_{\D_\s}(h)-4\sum_{i=\e,T} H_{h_i-1}\tilde P_i\, S_{-\D_\s+\t_i/2}(h)
\,,
\label{P0-IeT}
\ee
where $\tilde P_i=\frac{\G(2h_i)}{\G(h_i)^2} P_i$ and the anomalous dimension is defined as 
$\g_i\equiv\t_i-2\D_\s$. 
The subscript $\{I,\e,T\}$ indicates the asymptotic behavior is associated with the cross-channel contributions of these three low-lying operators. 
As before, we can resum the GFF contributions.  
The nonperturbative crossing constraints can be expressed as  
the difference between the contributions involving the free parameters and their GFF counterparts: 
\be
\mathcal G_{\{I,\e,T\}}^{(m,n)}
&\equiv&\sum_{i=\e,T} P_i\, \mathcal F^{(m,n)}_{\D_i,\ell_i}
+\sum_{\ell=4,\dots,\ell^\ast}P_{\{I,\e,T\}}^{(k=0)}(h)\, \mathcal F^{(m,n)}_{2h-\ell,\ell}
+\sum_{\ell=0,2}P_{\{I\}}^{(k=1)}\Big(\frac{\D+\ell}{2}\Big)\, \mathcal F^{(m,n)}_{\D,\ell}\Big|_{\D=\D^{(k=1)}_\ell}
\nn&&
-\sum_{\ell=0,2,\dots,\ell^\ast}P_{\{I\}}^{(k=0)}(\D_\s+\ell)\, \mathcal F^{(m,n)}_{2\D_\s+\ell,\ell}
-\sum_{\ell=0,2}P_{\{I\}}^{(k=1)}\Big(\frac{\D+\ell}{2}\Big)\, \mathcal F^{(m,n)}_{\D,\ell}\Big|_{\D=2\D_\s+2+\ell}
\nn&\approx& 0\,,
\label{crossing-IeT}
\ee
where the conformal spin $h$ in the second summation is approximated by
\be
h=\D_\s+\ell+\frac 1 2 \g_{\{I,\e,T\}}^{(k=0)}(\D_\s+\ell)\,.
\label{conformal-spin-approximation}
\ee 
When $\ell^\ast$ is sufficiently large, the results are not sensitive to this parameter. 
\footnote{We find that $\ell^\ast=10$ is large enough in our discussion. }
Note that the free parameters $\D^{(k=1)}_{\ell=0,2}$ are the scaling dimensions of the second lowest $\mathbb Z_2$-even operators with spin $\ell=0,2$. 
As explained above, their OPE coefficients are still approximated by the GFF formula \eqref{P-I-1}.  

Since there are 6 free parameters, 
we consider the following set of nonperturbative crossing constraints
\be
(m,n)=(0,1),(0,3),(1,2), (0,5), (1,4), (2,3)\,,
\label{derivative-IeT}
\ee
where the total number of derivatives is at most 5. 
As the number of free parameters and constraints match, 
one may want to impose $\mathcal G_{\{I,\e,T\}}^{(m,n)}=0$ with \eqref{derivative-IeT}. 
However, this will rule out the Ising solution, and we only find the free theory solution. 
\footnote{We do not find a solution satisfying all the nonperturbative constraints around $(\D_\s,\D_\e)=(0.52, 1.4)$. } 
As a result, we should not impose that $\mathcal G_{\{I,\e,T\}}^{(m,n)}$ vanish exactly. 
To obtain the 3D Ising solution, we need to consider a weaker formulation of the crossing constraints \eqref{crossing-IeT}. 

A natural modification is to search for the approximate solutions with small errors in the crossing constraints \eqref{crossing-IeT}. 
To measure the errors of the crossing constraints, 
we introduce the $\eta$ function:
\be
\eta=\sqrt{\sum_{m,n}\Big|\frac{1}{m!n!}\mathcal G_{\{I,\e,T\}}^{(m,n)}\Big|^2}\,,
\label{eta-function-definition}
\ee
where we use $(m!n!)^{-2}$  to suppress the high derivative constraints 
as they converge slower and are relatively less sensitive to the low-lying parameters. 
Instead of imposing the exact constraints $\mathcal G_{\{I,\e,T\}}^{(m,n)}=0$, 
we now search for the local minima of the $\eta$ function. 
In \cite{Li:2017ukc}, the error minimization approach was proposed as 
a potentially more systematic formulation of the truncation approach initiated by Gliozzi \cite{Gliozzi:2013ysa}. 
\footnote{See \cite{Gliozzi:2014jsa,Gliozzi:2015qsa,Nakayama:2016cim,Gliozzi:2016cmg,Esterlis:2016psv,Hasegawa:2016piv,Hikami:2017hwv,Hikami:2017sbg,Hikami:2018mrf,Leclair:2018trn,Hikami:2018qpz,Rong:2020gbi,Nakayama:2021zcr,Padayasi:2021sik} for the use of Gliozzi's determinant method or the singular value method in the conformal bootstrap studies. }
In fact, the minimization formulation of \cite{Li:2017ukc} was partly motivated by 
absence of solutions to some crossing symmetric ansatz in the inverse bootstrap approach \cite{Li:2017agi}, 
which is similar to the situation here.
The error minimization approach has been adopted  
in a number of bootstrap studies \cite{Iliesiu:2018zlz,Li:2021uki,Kantor:2021kbx,Kantor:2021jpz,Li:2022prn,Laio:2022ayq,Kantor:2022epi,Poland:2023vpn,Niarchos:2023lot}. 
\footnote{The minimization can also be implemented with stochastic algorithms \cite{Kantor:2021kbx,Kantor:2021jpz,Laio:2022ayq,Kantor:2022epi,Niarchos:2023lot}. 
In the recent work \cite{Niarchos:2023lot}, 
the determinations of the OPE coefficients from the truncation method were improved by 
using 10 exact scaling dimensions from the integrability method and 
introducing 52 effective operators to approximate the high-lying contributions,  
where the scaling dimensions of the effective operators are free parameters 
and are determined by the stochastic minimization under some positivity constraints. 
In this work, we do not use any external input scaling dimensions nor positivity constraints.  
} 

\begin{figure}[h]
	\centering
		\includegraphics[width=1\linewidth]{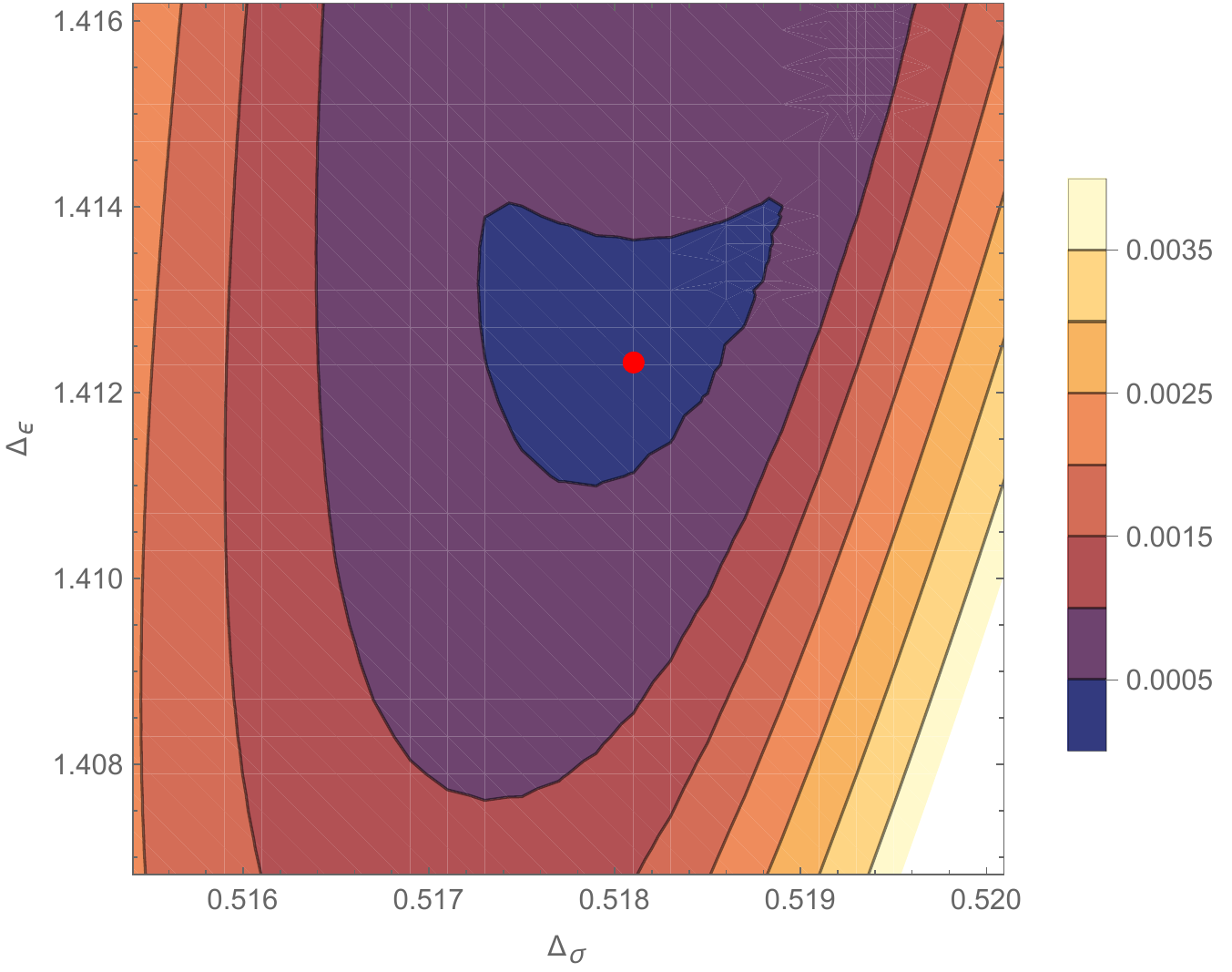}
	\caption{The minimized $\eta$ function \eqref{eta-function-definition} at different $(\D_\s, \D_\e)$. 
	The local minimum (red dot) with $\eta_\text{min}\approx 0.00043$ determines the 3D Ising results in \eqref{sol-IeT}. }
	\label{Fig-eta}
\end{figure}

As expected for the 3D Ising CFT, 
there exists a local minimum of the $\eta$ function around $(\D_\s, \D_\e)\approx (0.518, 1.41)$, 
which is indicated by the red point in Fig. \ref{Fig-eta}. 
The resulting low-lying parameters are 
\footnote{We use Mathematica's FindMinimum. The working precision is set to 200. }
\be
\D_\s\approx 0.51810\,,\quad
\D_\e\approx 1.4123\,,\quad
P_\e\approx 1.1053\,,\quad
P_T\approx 0.10650\,, 
\label{sol-IeT}
\ee
where only 5 significant digits are presented. 
For comparison, the accuracy of \eqref{sol-IeT} is better than the first positive bootstrap results for the 3D Ising model in \cite{El-Showk:2012cjh}, 
even though the nonperturbative constraints in \cite{El-Showk:2012cjh} involve 23 derivatives
and the number of constraints was significantly greater than our case with at most 5 derivatives. 
Using the positivity bound results as reference values, 
the relative errors of the $\eta$ minimization results are $(-0.009\%,\,  -0.02 \%,\,  -0.1\%,\, 0.1\%)$. 

We can also estimate the errors without using the positivity bound results.  
Let us substitute the free parameters with the $\eta$ minimization results in \eqref{sol-IeT},  
except for the one under error estimation,  
and then solve for this parameter using one of the crossing constraints in \eqref{derivative-IeT}. 
The largest deviation from the $\eta$ minimization result in \eqref{sol-IeT} gives an error estimate of this parameter \cite{Li:2017ukc}.  
The results are 
$|\d\D_\s|=0.00145\,,
|\d\D_\e|=0.0007\,,
|\d P_\e|=0.0009\,,
|\d P_T|=0.00039$.

The corresponding critical exponents are
\be
\eta\approx 0.03620\,,\quad\nu\approx 0.62984\,,
\ee
where the relative errors of the $\eta$ minimization results are $(-0.3\%,-0.02\%)$. 
These results are more accurate than those in Sec. \ref{Sec_I}. 
The estimated errors without using the positivity bound results are
$|\d\eta|=0.00290\,,|\d\nu|=0.00030$. 
We also obtain the scaling dimensions of the low spin operators in the higher trajectory 
$(\D_{\ell=0}^{(k=1)},\D_{\ell=2}^{(k=1)})\approx (3.68, 6.29)$, 
which are less accurate. 
For comparison, the  estimates for $(\D_{\ell=0}^{(k=1)},\D_{\ell=2}^{(k=1)})$ in \cite{Simmons-Duffin:2016wlq} are
$(3.82968(23), 5.50915(44))$. 

\begin{figure}[h]
	\centering
		\includegraphics[width=1\linewidth]{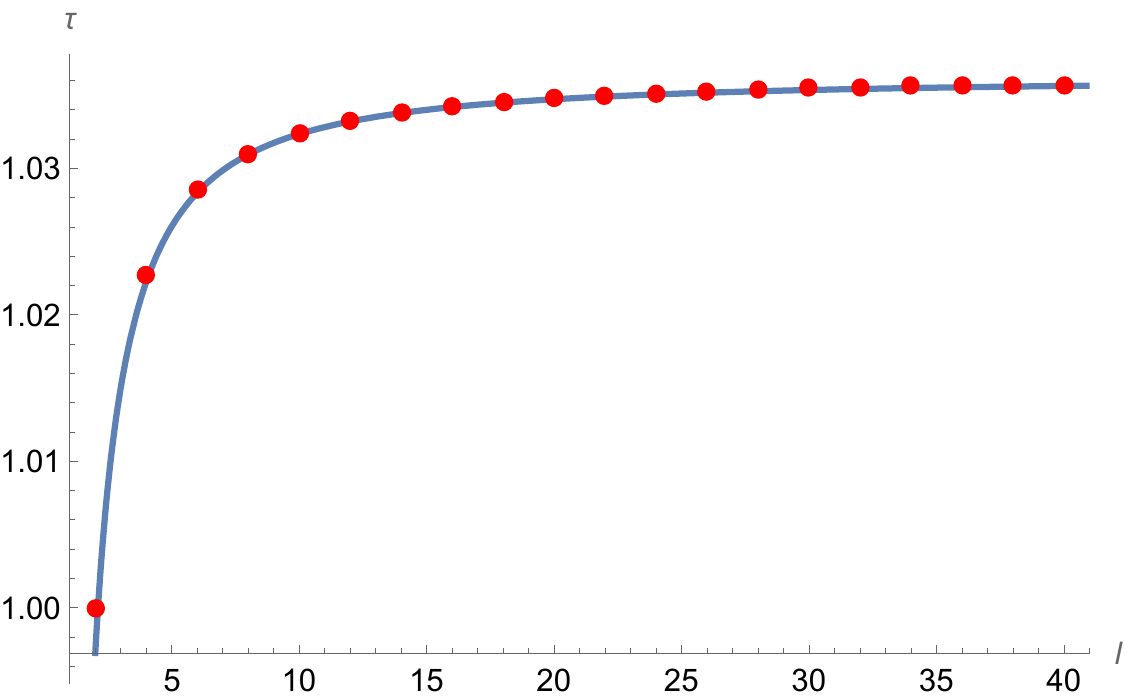}
	\caption{The twist $\t=\D-\ell$ of the lowest $\mathbb Z_2$-even trajectory at various spin $\ell$. 
	The curve is based on the analytic expression $\t(\ell)=2\D_\s+\g_{\{I,\e,T\}}^{(k=0)}(\D_\s+\ell)$. 
	The red dots are the numerical estimates from \cite{Simmons-Duffin:2016wlq}. 
}
	\label{trajectory-twist-plot}
\end{figure}

\begin{figure}[h]
	\centering
		\includegraphics[width=1\linewidth]{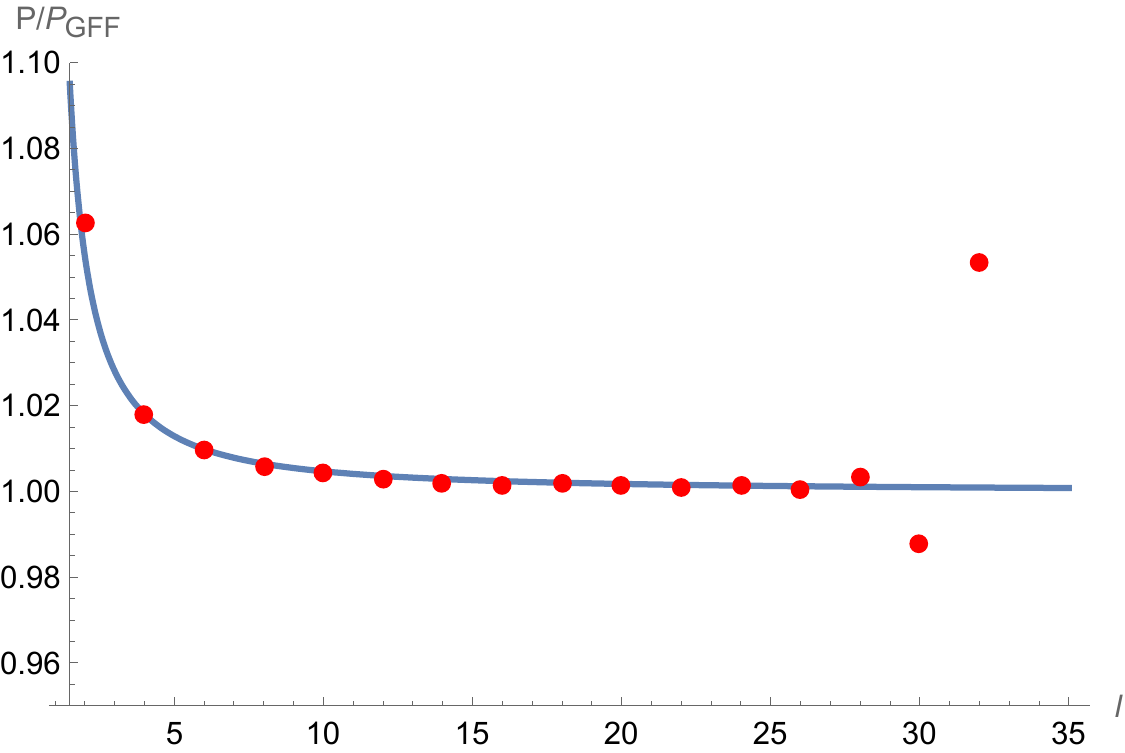}
	\caption{The squared OPE coefficients of the leading $\mathbb Z_2$-even trajectory at different spin $\ell$, 
	divided by the generalized free case $P_\text{GFF}(\ell)=2S_{-\D_\s}(\D_\s+\ell)$.  
	The curve is based on the analytic expression $P(\ell)=P_{\{I,\e,T\}}^{(k=0)}(h)$ in \eqref{P0-IeT}, where $h$ is approximated by \eqref{conformal-spin-approximation}. 
	The red dots are the numerical estimates from \cite{Simmons-Duffin:2016wlq}, 
	which have noticeable errors at $\ell>26$. 
}
	\label{trajectory-OPE-plot}
\end{figure}

Using the precise bootstrap bound results as the input parameters for the analytic formulae, 
one can compare the analytic asymptotic formulae with 
the numerical estimates \cite{Alday:2015ota, Alday:2015ewa, Simmons-Duffin:2016wlq}, 
whose agreement is surprisingly good and motivated the present work. 
For completeness, we also present the comparisons in 
Fig. \ref{trajectory-twist-plot} and Fig. \ref{trajectory-OPE-plot}, 
whose agreement with the numerical estimates \cite{Simmons-Duffin:2016wlq} is excellent for $\ell\geq 4$. 
The main difference from the earlier studies \cite{Alday:2015ota, Alday:2015ewa, Simmons-Duffin:2016wlq} is that 
the input parameters $(\D_\s,\D_\e,P_\e,P_T)$ in \eqref{sol-IeT} are not extracted from the bootstrap bound results. 
As a step forward, both the low-lying scaling dimensions and OPE coefficients are determined by the hybrid numerical/analytic approach and the $\eta$ minimization,  
without relying on the results from other approaches.  
\footnote{Since the conformal blocks are highly nonlinear functions in the scaling dimensions, 
it is more difficult to determine the low-lying scaling dimensions than the coefficients of the conformal blocks. 
In \cite{Iliesiu:2018zlz, Poland:2023bny,Niarchos:2023lot}, 
the low-lying scaling dimensions are input parameters from the positivity bounds or integrability method  
and the improved truncation methods are mainly used to determine the coefficients of conformal blocks. 
 }
 
After taking into account the leading corrections to the GFF behavior, 
our results also become much more stable under a change in the selection of derivatives, 
in comparison to those in Sec. \ref{Sec_I}. 
If we substitute the $(m,n)=(0,5)$ constraint with a 7 derivative constraint, 
the results are less accurate than the results \eqref{sol-IeT} with at most 5 derivatives, 
but still better than the results \eqref{sol-GFF} based on the GFF approximations. 
The corresponding $\eta$ minimization results are
\be
(0,7):\quad
\D_\s\approx 0.51781\,,\quad
\D_\e\approx 1.4122\,,\quad
P_\e\approx 1.1042\,,\quad
P_T\approx 0.10642\,,
\label{0-7}
\ee
\be
(1,6):\quad
\D_\s\approx 0.51867\,,\quad
\D_\e\approx 1.4133\,,\quad
P_\e\approx 1.1069\,,\quad
P_T\approx 0.10658\,,
\label{1-6}
\ee
\be
(2,5):\quad
\D_\s\approx 0.51876\,,\quad
\D_\e\approx 1.4137\,,\quad
P_\e\approx 1.1069\,,\quad
P_T\approx 0.10659\,,
\label{2-5}
\ee
\be
(3,4):\quad
\D_\s\approx 0.51862\,,\quad
\D_\e\approx 1.4131\,,\quad
P_\e\approx 1.1068\,,\quad
P_T\approx 0.10657\,,
\label{3-4}
\ee
where the local minimum of the $\eta$ function is zero expect for the first case. 
If we replace the $(1,4)$ or $(2,3)$ constraint with a 7-derivative one, 
we still find a local minimum of the $\eta$ function around the 3D Ising solution, 
but the accuracy is lower than \eqref{sol-GFF}. 
It seems that the crossing constraints with $m\sim n$ are more sensitive to the low-lying contributions 
than those with $m\ll n$, so the former can lead to more accurate determinations for the low-lying parameters. 
 
We can also add more constraints to the $\eta$ functions, 
leading to an overdetermined system with $\eta_\text{min}>0$. 
If we use all the crossing constraints with $m+n\leq 7$, 
the local minimum of the $\eta$ function around $(\D_\s, \D_\e)=(0.52,1.41)$ gives
$\D_\s\approx 5.1974\,,
\D_\e\approx 1.4192\,,
P_\e\approx 1.1019\,,
P_T\approx 0.10715$
and
$(\D_{\ell=0}^{(k=1)},\D_{\ell=2}^{(k=1)})\approx (3.75, 5.38)$. 
Although  less accurate in comparison to \eqref{sol-IeT}, 
the $\eta$ minimization results for the relevant operators 
are still around the 3D Ising solution. 
On the other hand, the irrelevant scaling dimensions $(\D_{\ell=0}^{(k=1)},\D_{\ell=2}^{(k=1)})$ become more accurate. 
\footnote{We also notice this improvement for the irrelevant scaling dimensions by replacing a 5-derivative constraint with a 7-derivative constraint. 
For example, the cases for \eqref{0-7}, \eqref{1-6}, \eqref{2-5}, \eqref{3-4} are 
$(\D_{\ell=0}^{(k=1)},\D_{\ell=2}^{(k=1)})\approx (3.75, 6.87), (3.74, 5.36), (3.77, 5.22),(3.72, 5.44)$. 
} 
This is consistent with our expectation that 
the higher derivative equations are more sensitive to the high-lying contributions, 
so the predictions for the irrelevant operators are improved, 
at the price of lowering the accuracy of the low-lying parameters, 
which are contaminated by the less accurate high-lying contributions. 

Since we have 10 constraints with $m+n\leq 7$, 
we can introduce more free parameters. 
It is natural to consider the lowest spin-4 operator, 
as its scaling dimension $\D_{\ell=4}^{(k=0)}\approx 5$ is close to 
the subleading spin-2 scaling dimension, which is viewed as a free parameter in the above.  
Surprisingly, we arrive at even more accurate results by relaxing the scaling dimension and OPE coefficient of the lowest spin-4 operator:
\be
\D_\s\approx 0.518152\,,\quad
\D_\e\approx 1.41278\,,\quad
P_\e\approx 1.10640\,,\quad
P_T\approx 0.106226\,.
\label{sol-IeTl4}
\ee
The spin-4 results are $\D^{(k=0)}_{\ell=4}\approx 5.021$ and $P^{(k=0)}_{\ell=4}\approx0.00479$.  
The irrelevant scaling dimensions are
$(\D_{\ell=0}^{(k=1)},\D_{\ell=2}^{(k=1)})\approx (3.67, 5.46)$. 
These are the most accurate results in this work. 
In particular, the OPE coefficient $\l_{\s\s\e}=P_\e^{1/2}\approx1.05186$ is significantly improved.  
The local minimization is carried out by 
Mathematica's FindMinimum, 
starting at $(\D_\s, \D_\e)=(1,3/2)$, 
$(\D_{\ell=0}^{(k=1)},\D_{\ell=2}^{(k=1)}, \D^{(k=0)}_{\ell=4})=(3,5,5)$ and 
$(P_\e,P_T,P^{(k=0)}_{\ell=4})=(1,1/10,1/200)$, 
which are some crude numbers from the GFF approximation.  
The corresponding critical exponents
\be
\eta\approx 0.036304\,,\quad\nu\approx 0.63003\,
\ee
are almost as accurate as the latest Monte-Carlo determinations \cite{Hasenbusch:2021tei}. 

As the system is still overdetermined, one may try to introduce more free parameters at higher spin. 
However, if we do not approximate the lowest spin-6 parameters by the analytic formulae 
and set them as free parameters, 
we are not able to find a local minimum around the 3D Ising solution. 
It seems that the $\eta$ minimization becomes unstable 
due to the fact that the squared OPE coefficient of the spin-6 operator is allowed to be negative.  
Besides reducing computational efforts, 
stabilization is another important merit of analytic approximations for the high-lying contributions.  
This is different from the hybrid approach in \cite{Su:2022xnj}, 
in which the use of positivity constraints stabilizes the hybrid bootstrap method  
and thus the effects of the choice for the gluing spin can be studied more systematically. 

\section{Discussion} 
\label{Sec_discussion}
In this work, we revisited the conformal bootstrap approach to the 3D Ising model
using the global asymptotic behavior and the rapidly convergent crossing constraints. 
We derived surprisingly accurate results \eqref{sol-IeTl4} 
from simple analytic asymptotic formulae 
and a small number of nonperturbative crossing constraints, 
using the error minimization method. 
In some sense, 
this furnishes a minimal example for the nonperturbative conformal bootstrap in three dimensions, with the help of an analytic understanding from the lightcone singularity structure. 

To improve the accuracy of the hybrid approach, 
it is crucial to go beyond the leading corrections to the GFF behavior. 
\footnote{Some examples of the subleading corrections are the Jacobian factors in the OPE coefficients and  
the higher spin contributions from the lowest twist trajectory. } 
The asymptotic formulae can be computed more systematically using the methods in 
\cite{Alday:2016njk,Alday:2016jfr,Simmons-Duffin:2016wlq}. 
However, one should be more careful as a higher order asymptotic expansion may give worse results at low spin, even if the approximations at large spin are improved. 
Then the corresponding low spin parameters should be viewed as free parameters 
and determined by the nonperturbative crossing constraints. 
To avoid this subtlety, one may use the convergent Lorentzian inversion formula 
\cite{Caron-Huot:2017vep,Simmons-Duffin:2017nub,Kravchuk:2018htv}, 
which includes the nonperturbative contributions 
\cite{Liu:2018jhs, Cardona:2018qrt, Albayrak:2019gnz, Li:2019dix, Albayrak:2020rxh, Liu:2020tpf, Caron-Huot:2020ouj,Atanasov:2022bpi}. 
\footnote{The choice of $z$ may be determined by the nonperturbative crossing constraints as the other free parameters.}
\footnote{A more rigorous alternative approach is to use the CFT dispersive sum rules
\cite{Caron-Huot:2020adz}. 
See also \cite{Carmi:2019cub,Mazac:2019shk,Penedones:2019tng,Carmi:2020ekr,Trinh:2021mll}.} 
However, the price to pay is a higher computational cost. 
In practice, it may be useful to combine the asymptotic methods with the convergent  methods.

We also need to unmix the higher trajectories. 
\footnote{The high-lying spectrum of the 2D Ising CFT is highly degenerate. 
One can derive the exact solutions  
without unmixing the high-lying contributions because the degenerate scaling dimensions are exactly the same. 
In this way, the averaged behavior encodes the exact information about the total contributions.  }
In the present work, we used the decoupling phenomenon to circumvent the mixing problem, 
and carried out the bootstrap study using only the identical correlator 
$\langle\s(x_1)\s(x_2)\s(x_3)\s(x_4)\rangle$. 
To properly resolve the mixing problem, one needs to consider a mixed system of correlators \cite{Simmons-Duffin:2016wlq,Liu:2020tpf,Caron-Huot:2020ouj,Atanasov:2022bpi}. 
The large repulsion of the twist spectra at low spin is particularly interesting.  
This can lead to more significant deviations from the leading GFF behavior. 
We also believe that this is closely related to the decoupling phenomenon in \cite{El-Showk:2012cjh,El-Showk:2014dwa}.

We would like to emphasize that the goal of the present work is not to surpass the remarkably precise results from the positivity bound approach.
\footnote{We need to take into account the subleading effects in a systematic and stable way, which is still not completely clear at this point, but we will try to achieve this in the future. } 
Instead, we want to show that a hybrid analytical/numerical  approach can have the great potential 
in reducing the computational cost  
with the help of a global understanding of the high-lying data. 
The application to the 3D Ising model is just the first example.  
This strategy can be applied to other well constrained bootstrap solutions, 
such as the O$(N)$ models \cite{Kos:2015mba,Kos:2016ysd,Chester:2019ifh,Chester:2020iyt} and 
the supersymmetric Ising model \cite{Rong:2018okz,Atanasov:2018kqw,Atanasov:2022bpi}. 
For a more complete list of the conformal bootstrap targets, we refer to the comprehensive review \cite{Poland:2018epd} and the more recent one \cite{Rychkov:2023wsd}. 
As we are not using any positivity constraint, 
a natural and important direction is to consider the less explored non-positive bootstrap problems 
that involve $D>2$ conformal field theories with nonunitarity \cite{Gliozzi:2013ysa}, 
defects \cite{Liendo:2012hy,Billo:2016cpy}, and on nontrivial manifolds \cite{Nakayama:2016cim,Iliesiu:2018fao}.

\section*{Acknowledgments}
I would like to thank Junchen Rong and Ning Su for discussions. 
I also thank the referee for many constructive suggestions. 
This work was supported by 
the Natural Science Foundation of China (Grant No. 12205386) and the Guangzhou Municipal
Science and Technology Project (Grant No. 2023A04J0006).

\end{document}